\documentclass[aps,pra,amsmath,amsfonts,floatfix,superscriptaddress,notitlepage,nofootinbib]{revtex4}
\usepackage{graphicx,graphics}
\usepackage[T1]{fontenc}
\usepackage{epstopdf}
\usepackage{amsmath}
\usepackage{amssymb}
\usepackage[colorlinks=true]{hyperref}

\usepackage{etoolbox}
\preto{\section}{\setcounter{subsubsection}{0}}

\let\a=\alpha \let\be=\beta  \let\g=\gamma  \let\d=\delta

   \let\th=\theta

\def\ie{{i.e. }}\def\eg{{e.g. }}

    \def\bJ{\boldsymbol{J}} 
 \def\bphi{\boldsymbol{\phi}} \def\bvphi{\boldsymbol{\varphi}}

\renewcommand\b[1]{\boldsymbol{#1}}

\def\to{\rightarrow}

\def\Tr{\text{Tr}}

\begin{document}

\title{On the effective action in presence of local non-linear constraints}
\author{Adam Ran\c con}
\address{Universit\'e de Lille, CNRS, UMR 8523 -- PhLAM -- Laboratoire de Physique des Lasers Atomes et Mol\'ecules, F-59000 Lille, France}
 \author{Ivan Balog}
\address{Institute  of  Physics,  Bijeni\v cka  cesta  46,  HR-10001  Zagreb,  Croatia}

\begin{abstract}
The conditions for the existence of the effective action in statistical field theory,  the Legendre transform of the cumulant generating function,  in presence of non-linear local constraints are discussed. This problem is of  importance for non-perturbative approaches, such as the functional renormalization group. We show that the Legendre transform exists as long as the non-linear constraints do not imply linear constraints on the microscopic fields. We discuss how to handle the case of effectively linear constraints and we naturally obtain that the second derivative of the effective action is the Moore-Penrose pseudo-inverse of the correlation function. We illustrate our discussion with toy-models, and show that the correct counting of degrees of freedom in non-linearly constrained statistical field theories can be rather counter-intuitive.
\end{abstract}
\maketitle

\section{Introduction}

Counting the number of degrees of freedom (DOF) in quantum and statistical field theory is a subtle issue, best exemplified by gauge theories. In (quantum) electrodynamics, gauge invariance leads to a reduced number of physical DOF, from four to two in four dimensions. In contrast, most models of statistical physics are defined in terms of microscopic DOF which satisfy hard local non-linear constraints, \ie the components of the field have to satisfy non-linear equations. One well-known example is the Ising model, which is defined in terms of classical spins on a lattice, each spin satisfying the constraint $s^2=1$. 
In the case of spin models, counting the number of DOF is somewhat intuitive: the number of degrees of freedom (per site) is given by the number of components of the spin --equivalent to the number of components of the $N$-component vector $\bvphi$, irrespective of the hard constraint $\bvphi^2=1$. There are however situations where the microscopic field has no intuitive physical meaning, as for instance in the case of  the replica field theory describing the Anderson transition, where the field is a matrix which squares to the identity matrix, see \eg \cite{Pruisken1982}. Then, it is much harder to gain intuition on the number of DOF (not even withstanding the issue of the zero replica limit).
When one is interested in the physics close to a second order phase transition, universality usually allows for relaxing the constraints and for working with a more manageable $\bvphi^4$ theory \cite{Zinn-JustinBook}.  However, a lot of information about the non-universal physics is lost in the process, which prevents from computing valuable nonuniversal quantities characterizing a system, such as the critical temperature. Taking into account the non-linear constraints as exactly as possible calls for functional methods, among which the Function Renormalization Group (FRG) is very well suited for the study of critical phenomena \cite{Berges2002,DelamotteIntro}.

In all generality, let us assume that the microscopic field (which is integrated over) is described in terms of $N$ real numbers (for instance, $\bvphi$ could be a vector with $N$ components, or a matrix with $N$ entries, etc.). Without constraints, it has $N$ ``linear degrees of freedom''. If we now introduce $M$ constraints, $\bvphi$ can be parametrized in terms of $N-M$ coordinates $x_{\alpha=1,\ldots,N-M}$ which we call the ``non-linear degrees of freedom'' (there can be additional discrete degrees of freedom, \eg the sign $s=\pm$ in the Ising case). Equivalently, we could choose $N-M$ components of $\bvphi$, consider them independent, and reexpress the $M$ components left in terms of these. 
Already, it is not clear which of the linear DOF or non-linear DOF should be considered as the ``real'' ones.

Furthermore, the microscopic field $\bvphi$ is not necessarily an observable, whereas its average, the ``magnetization'' $\bphi=\langle \bvphi\rangle$, usually is. Should we consider instead its DOF as the true ones? Does $\bphi$ even have the same number of linear or non-linear DOF than $\bvphi$? These are important issues, since, for instance, in a modern formulation of the FRG \cite{Berges2002}, one works with the effective action $\Gamma[\bphi]$, which is a functional of $\bphi$. It is defined as the Legendre transform of the logarithm of the generating function $Z[\bJ]$, where $\bJ$ is a $N$-component source linearly coupled to  $\bvphi$, a transformation which involves expressing the source in terms of the magnetization.  While this construction can be done perturbatively in standard $\bvphi^4$ theories  \cite{Zinn-JustinBook}, it is far from obvious that this is possible when there are non-linear constraints, especially if the number of non-linear constraints $M$ is of the order of $N$.  Indeed, one could imagine that these constraints on $\bvphi$ also imply constraints on $\bphi$ of some sort (reducing the number of independent components of $\bphi$), and thus rendering the Legendre transform ill-defined (since the source $\bJ$ would have more independent components than $\bphi$).  Understanding how to choose the source in order to obtain a well-defined effective action is thus of great importance for the applicability of the FRG to models with non-linear constraints.
 
To show how this problem can be very counter-intuitive, we can focus on a toy-model where spins are replaced by $SO(3)$ matrices. There, the microscopic fields, which are $3\times3$ real matrices $\bvphi\in SO(3)$, obey the non-linear constraints $\bvphi.\bvphi^T=\mathbf{1}$ and $\det\bvphi=1$, and can be parametrized by only three Euler angles, the corresponding non-linear DOF. How many ``true'' DOF does this field theory have? How many independent DOF does the magnetization $\bphi$ have, and how should the source be chosen to properly define the effective action? It so happens that the naive answers (three DOF for the three Euler angles, or maybe four DOF for the three angles plus an ``amplitude'') are completely wrong: the true answer is nine DOF, and the magnetization is oblivious of the six non-linear constraints imposed on the microscopic field.   
 
It is the goal of this paper to clarify these points. The main result is that the ``true'' DOF in this context are the ``linear degrees of freedom'', and more precisely, the number of linearly independent components of the microscopic field. Obviously, linear constraints on $\bvphi$ (\eg $\bvphi=\bvphi^T$ if $\bvphi$ is a symmetric matrix) do reduce the number of DOF, and the effective action is defined only in terms of a magnetization that respects the linear constraint (\ie $\bphi=\bphi^T$ in this example). On the other hand, non-linear constraints that \emph{do not imply} linear constraints do not change the number of linear DOF, and in particular, $\bphi$ has in practice $N$ independent components. It is as if the non-linear constraints are washed out by the thermal averaging, and the averaged fields are in essence insensitive to the constraints. This is the explanation for the $SO(3)$ matrix toy-model.
While it is not surprising for spin systems, where the magnetization is a $N$-component vector (of maximum length one), we will exemplify this on the rather counter-intuitive $SO(n)$ toy-models  where the number of constraints is larger than the number of non-linear DOF (\ie $M>N/2$), and still the magnetization has $N$ independent components.
Finally, there is the intermediate case, where the $M$ non-linear constraints \emph{do imply} $m$ linear constraints. One then finds that $\bphi$ does respect these $m$ linear constraints, and taking them into account insures the existence of the effective action. The remaining $M-m$ non-linear constraints do not pose any additional problem. Stated otherwise, only purely linear constraints (possibly induced by the non-linear ones) prevent one to perform the Legendre transform naively, and need a special care.


We will illustrate our discussions with toy-models, that will exemplify different cases of constraints discussed above, and which are introduced in Sec.~\ref{sec_toymodels}. Sec.~\ref{sec_existenceTL} gives a general discussion on how to count the DOF and the conditions for the existence of the effective action. In Sec.~\ref{sec_derivative} we present how a definition of constrained derivatives can be used in the case of (effectively) linear constraints, to work as if there were none. Finally, we discuss our findings and open problems in Sec.~\ref{sec_concl}.

\section{Toy-models \label{sec_toymodels}}

   We introduce single-site toy-models that illustrate one of the cases of linear or non-linear constraints discussed in the introduction. For a model with purely linear constraints, we use a model of $2\times2$ symmetric matrices. In the case of purely non-linear constraints, we use the XY model (planar spins of unit length), on which we can base our intuition. Finally, we introduce models of matrices  belonging to $SO(n)$. In particular, the $SO(3)$ has purely non-linear constraints,  with $N=9$ linear DOF and $M=6$ constraints, while the $SO(2)$ model has three non-linear constraints, which in fact imply effectively two linear constraints and one non-linear.

\subsubsection{Symmetric matrices \label{subsec_sym}}

Consider first the case of $2\times2$ real symmetric matrices. It is relevant because it is straightforward to generalize the discussion below to arbitrary symmetric matrices. A simple model of real $n\times n$ symmetric matrices is the Gaussian Orthogonal Ensemble of Random Matrix Theory $GOE(n)$. 

The microscopic field $\bvphi$ can be parametrized as


\begin{equation}
\bvphi=\begin{pmatrix}
a & c \\
c & b
\end{pmatrix},
\label{eq_sym_vphi}
\end{equation}
with $a,b,c$ real numbers, where the hard linear constraint $\bvphi=\bvphi^T$ is imposed by construction. The partition function is given by
\begin{equation}
\begin{split}
Z[\bJ]=&\int d\bvphi e^{-\frac12 \Tr(\bvphi^2)+\Tr(\bJ \bvphi)},\\
=&\int\frac{da\, db\, dc}{2\pi^{3/2}}e^{-\frac12 \Tr(\bvphi^2)+\Tr(\bJ \bvphi)},
\end{split}
\end{equation}
which is invariant under $\bvphi\to \b V \bvphi \b V^T$, with $\b V\in O(2)$. An explicit calculation of the generating function $W[\bJ]=\ln Z[\bJ]$ for an arbitrary source gives
\begin{equation}
W[\bJ]=\frac14(\Tr(\bJ^2)+\Tr(\bJ \bJ^T)).
\label{eq_defW_sym}
\end{equation}
Note that, in addition to being invariant under $\bJ\to \b V^T \bJ \b V$, $W[\bJ]$ does not depend on the antisymmetric part of $\bJ$ due to the constraint, which is a first hint that the Legendre transform will not exist if performed naively.

In particular, the magnetization is given by
\begin{equation}
\bphi[\bJ]=\frac{\bJ+\bJ^T}{2},
\end{equation}
and is explicitly symmetric for arbitrary sources. Clearly, if $\bJ$ is not symmetric, it has four independent components, while the magnetization has only three. The relationship between source and magnetization cannot be inverted, preventing the naive definition of the effective action.

\subsubsection{$XY$ model \label{sssec_XY}}
One of the simplest models with purely non-linear constraints is that of a planar classical spin of unit length, $\bvphi=(\varphi_1,\varphi_2)$  with the constraint $C(\bvphi)=\varphi_1^2+\varphi_2^2-1=0$. Obviously, $\bvphi$ can be parametrized by one angle $\theta$, $\bvphi=(\cos\theta,\sin\theta)$. Introducing a magnetic field (a source) $\bJ=(J_1,J_2)$, the partition function is given by
\begin{equation}
\begin{split}
Z[\bJ]&=\int_0^{2\pi} \frac{d\theta}{2\pi}e^{J_1\cos\theta +J_2\sin\theta},\\
&=I_0\left(J\right),
\end{split}
\end{equation}
with $J=|\bJ|$ and $I_n(z)$ the $n$-th modified Bessel function. The magnetization is easily obtained 
\begin{equation}
\bphi[\bJ]=\frac{\bJ}{J}\frac{I_1(J)}{I_0(J)}.
\end{equation}
Of course, the magnetization points in the direction of the source, while its amplitude is bounded from above, $\phi[\bJ]\leq 1$. Furthermore, the relationship between $\bphi$ and $\bJ$ can be inverted, pointing toward a simple definition of the effective action.

\subsubsection{$SO(n)$ matrix model}
Finally, we introduce matrix models with non-linear constraints. Here, we will focus on constraints imposing $\bvphi^T \bvphi=\mathbf{1}$ and $\det \bvphi=1$, where $\bvphi$ is a $n\times n$ matrix ($N=n^2$) belonging to the special orthogonal group $SO(n)$. Since $SO(n)$ has $\frac{n(n-1)}{2}$ generators, $\bvphi$ can be parametrized by as many angles. The integration measure is chosen to be the invariant Haar measure of the corresponding group. We will mostly study the two simplest cases $n=2$ and $n=3$, which corresponds to $N=4$ components with $M=3$ constraints, and $N=9$ and $M=6$, respectively. 
With these models, it is \emph{a priori} not clear whether these non-linear constraints will lead to problems for the definition of the effective action or not.

With the scalar product $\bJ.\bvphi={\rm tr}(\bJ^T\bvphi)$, $\bJ$ in the space of $n\times n$ real matrices $\mathcal{M}_n(\mathbb{R})$, and using the invariance of the measure, one shows that that the partition function is invariant under $\bJ\to \b{O}_1\bJ \b{O}_2$, $\b{O}_1, \b{O}_2\in SO(n)$. Thanks to Cayley-Hamilton theorem, it is a function of  $\det \bJ$ and ${\rm tr}\left((\bJ\bJ^T)^p\right)$, $1\leq p <n$.\footnote{Since $\b{O}_1, \b{O}_2\in SO(n)$ and not $O(n)$, the sign of $\det \bJ$ does matter, and one cannot use the invariants ${\rm tr}\left((\bJ\bJ^T)^p\right)$, $1\leq p \leq n$. } 

\paragraph*{$SO(2)$ matrix model-} We use the standard parametrization of the two-dimensional rotations with one angle $\th\in[0,2\pi]$,
\begin{equation}
\bvphi=\begin{pmatrix}
\cos(\th) & \sin(\th) \\
-\sin(\th) & \cos(\th)
\end{pmatrix},
\end{equation}
with invariant measure $d\mu(\bvphi)=\frac{d\th}{2\pi}$. Note that the three independent non-linear constraints give effectively rise to two linear ones, between the diagonal and off-diagonal elements. The partition function is 
\begin{equation}
\begin{split}
Z[\bJ]&= I_0\left(f(\bJ)\right),
\end{split}
\label{eq_WSO2}
\end{equation}
with $f(\bJ)=\sqrt{\Tr(\bJ\bJ^T)+2\det \bJ}=\sqrt{(J_{11}+J_{22})^2+(J_{12}-J_{21})^2}$, and the magnetization reads
\begin{equation}
\begin{split}
\phi_{11}[\bJ]=\phi_{22}[\bJ]=\frac{J_{11}+J_{22}}{f(\bJ)}\frac{I_1\left(f(\bJ)\right)}{I_0\left(f(\bJ)\right)},\\
\phi_{12}[\bJ]=-\phi_{21}[\bJ]=\frac{J_{12}-J_{21}}{f(\bJ)}\frac{I_1\left(f(\bJ)\right)}{I_0\left(f(\bJ)\right)}.
\end{split}
\end{equation}
The fact that $\phi_{11}[\bJ]=\phi_{22}[\bJ]$ and $\phi_{12}[\bJ]=-\phi_{21}[\bJ]$ for an arbitrary source $\bJ$ shows that it is impossible to invert the relationship between $\bphi$ and $\bJ$, precluding a naive Legendre transform for the $SO(2)$ matrix model.

\paragraph*{$SO(3)$ matrix model-}
Any matrix $\bvphi\in SO(3)$ can be parametrized by three rotations, thus three angles, with respect to two axis $x$ and $z$,
\begin{equation}
\bvphi=R_x(\chi) R_z(\theta) R_x(\xi),
\end{equation}
with $\chi,\xi \in [0,2\pi[$ and $\th \in [0,\pi[$, and with invariant measure 
\begin{equation}
d\mu(\bvphi)=\frac{\sin \th}{8\pi^2}d\th d\xi d\chi.
\end{equation}
Unfortunately, the partition function of the $SO(3)$ matrix model does not seem to have an explicit expression \cite{mathoverflow}. However, one can compute its small $\bJ$ expansion to find
\begin{equation}
\begin{split}
Z[\bJ]&=1+\frac{{\rm tr}\left(\bJ \bJ^T\right)}{6}+\frac{\det \bJ}{6}+\mathcal O (\bJ^4),\\
\end{split}
\end{equation}
while to lowest order in $\bJ$, the magnetization reads
\begin{equation}
\bphi[\bJ]=\frac{\bJ}{3}+\mathcal O (\bJ^2).
\end{equation}
At least for small sources, we observe that contrary to the $SO(2)$ matrix model, one can invert the relationship between $\bphi$ and $\bJ$, giving a hint that the effective action is well defined for $SO(3)$. This can in fact be generalized to all $n>2$ (see below).

\section{Effective action in presence of constraints \label{sec_existenceTL}}

   After giving concrete models in the previous section, we will now discuss the problems of defining the effective average action in a constrained systems using general terms. Our discussion will specifically address the case of single-site models, since the question of existence of the effective action is already present in the local limit, corresponding to only one field $\bvphi$ and its conjugated source $\bJ$. It is important to realize that the focus on the local case does not in any way restrict the applicability of our conclusions for systems of arbitrary dimensions.

\subsection{Notations and definitions}

In the absence of constraints, the microscopic field $\bvphi$ will belong to some vector space $E$ of dimension $d_E=N$, \eg $E=\mathbb{R}^N$ the space of real $N$-vectors, or $E=\mathcal{M}_n(\mathbb{R})$ the space of $n\times n=N$ real matrices. 
This space comes with a natural basis $\{\b{e}_A\}_{A=1,\ldots, N}$ and a scalar product $\b e_A.\b e_B=\delta_{AB}$. The microscopic field can be written as $\bvphi=\varphi_A \b e_A$, with $\bvphi=\{\varphi_A\}_{A=1,\ldots,N}$ the $N$ local microscopic degrees of freedom of our field theory (here $A$ can be a collective index, for example if $\bvphi$ is a matrix, and we use Einstein summation notation for repeated indices). These $\varphi_A$ corresponds to the linear degrees of freedom of the unconstrained field. The source $\bJ=J_A \b e_A$ is in principle of the same dimension as the unconstrained version of $\bvphi$, in order to generate all correlation functions of $\bvphi$.
Furthermore $\bvphi$ satisfies $M$ (linear or non-linear) independent constraints $C_a(\bvphi)=0$ ($a=1,\ldots,M$), which translate into $M$ relationships between the coefficients $\varphi_A$. 

The partition function is defined as
\begin{equation}
Z[\bJ]=\int d\mu(\bvphi) \,e^{ \bJ.\bvphi},
\label{eq_Zk}
\end{equation}
with $d\mu(\bvphi)$ the integration measure that we take of the form 
\begin{equation}
d \mu(\bvphi)=d\bvphi\, e^{-H(\bvphi)}\prod_{a=1}^M \delta\left(C_a(\bvphi)\right).
\end{equation}
Here $d\bvphi=\prod_A d\varphi_A$ is the flat measure on $E$ and $e^{-H(\bvphi)}$ is a positive weight that can for example be gaussian. In presence of $M$ (arbitrary) constraints, the allowed microscopic field configurations  belong to a set $S$ which is just the support of the measure $d\mu(\bvphi)$. Note that in principle, the scalar product in Eq.~\eqref{eq_Zk} could be replaced by a non-positive-definite symmetric bilinear form. This more general case is addressed in Appendix \ref{app_general}. Here we have $\bJ.\bvphi=J_A \varphi_A$, \eg $\bJ.\bvphi={\rm tr}(\bJ^T\bvphi)$ if  $\bJ$ and $\bvphi$ are real matrices.

Introducing the cumulant generating functional $W[\bJ]=\ln Z[\bJ]$, the order parameter, or magnetization, $\bphi[\bJ]$ is defined by 	
\begin{equation}
\phi_A[\bJ]\equiv \langle \varphi_A\rangle_{\bJ} =\frac{\delta W}{\delta J_A}[\bJ].
\end{equation}
The effective action $\Gamma[\bphi]$ is the Legendre transform of $W[\bJ]$, \ie it is given by
\begin{equation}
\begin{split}
\Gamma[\bphi]&=-W\left[\bJ[\bphi]\right]+ \bphi.\bJ[\bphi] ,\\
\frac{\delta\Gamma}{\delta \phi_A}&=J_A,
\end{split}
\end{equation}
where $\bJ[\bphi]$ is understood as the solution to $\frac{\delta W}{\delta J_A}\big|_{\bJ=\bJ[\bphi]}=\phi_A$ for $A=1,\ldots,N$. The Legendre transform of $W[\bJ]$ exists only if it is strictly convex, which allows for inverting unambiguously  $\bphi[\bJ]$ into $\bJ[\bphi]$. Convexity of $W[\bJ]$ is defined as $W[(1-t) \bJ_1+t \bJ_2]\leq (1-t)W[ \bJ_1]+tW[ \bJ_2]$, for all $\bJ_1$ and $\bJ_2\neq \bJ_1$, while strict convexity implies a strict inequality. \footnote{ Note that the strict convexity of $Z[\bJ]$ does not imply that of $W[\bJ]$: for this, $Z[\bJ]$ should be strictly log-convex.}

To find the conditions for strict convexity of $W[\bJ]$ in presence of constraints, it is useful to define a good basis in which to expand the constrained field, as well as the source.
However, since the constrained microscopic fields usually do not form a vector space (\eg the sum of two $XY$ spins is generically not an $XY$ spin), one cannot directly speak of the dimension of the constrained field, as one would do in the absence of constraints. This problem is easily avoided by using the space spanned by the microscopic field. While $S$ is generically not a vector space, its span $span(S)$ will be. By construction $span(S)\subseteq E$, and can be of smaller dimension than $E$, $dim( span(S))=N-m$, with $0\leq m<N$. Then, we have that $E=span(S)\oplus span(S)^\perp$, with $span(S)^\perp$ the orthogonal complement of $span(S)$. One can define for both spaces an orthogonal basis, $\{\b{f}^\parallel_i\}_{i=1,\ldots, N-m}$ and $\{\b{f}^\perp_i\}_{i=1,\ldots, m}$ for $span(S)$ and  $span(S)^\perp$ respectively.

 Because $S\subseteq span(S)$, we have $\b{f}^\perp_i. \bvphi=0$, $\forall \bvphi\in S$ and $i=1,\ldots,m$. This is equivalent to say that we have $m$ linear constraints on $\bvphi$. One easily shows that the converse is true, that is, having $m$ linear constraints of $\bvphi$ implies that $d_S\equiv dim( span(S))=N-m$. It is then natural to define the dimension of the constrained field as $dim( span(S))=N-m$. This is the number of linear DOF of $\bvphi$, which is in general different from the number $N-M$ of non-linear DOF.

A direct consequence of these results is that the derivatives of $W[\bJ]$ with respect to $\bJ$, with $\bJ\in E$, can be interpreted as symmetric multilinear forms with kernel $span(S)^\perp$. For example, $\b{e}_A\frac{\delta W}{\delta J_A}=\langle \bvphi\rangle_{\bJ} $ has obviously $span(S)^\perp$ as a kernel, and the same goes for $\frac{\delta^2 W}{\delta J_A\delta J_B}\b{e}_A\otimes\b{e}_B=\langle \bvphi\otimes\bvphi\rangle_{\bJ}$. An important consequence for the latter is that it is not invertible in a matrix sense if $span(S)^\perp$ is not empty.

\vspace{0.5cm}
\paragraph*{Application to the toy-models --}
For the $XY$ model, the vector space of the unconstrained field is $E=\mathbb{R}^2$, with dimension $d_E=2$. The set $S$ where the $XY$ spins live is the unit circle, the span of which is the whole two-dimensional plane, \ie $span(S)=E$. Therefore, while the number of non-linear DOF is $1$, the dimension of the XY spins is in fact $d_S=2$.

For both real symmetric $n\times n$ matrices and the $SO(n)$ matrix model, the space of the unconstrained fields is $E=\mathcal{M}_n(\mathbb{R})$ of dimension $d_E=n^2$. In the case of symmetric matrices, the linearly constrained space is still a vector space, of dimension $d_S=\frac{n(n-1)}{2}$. For $SO(2)$, the $m=2$ effectively linear constraints ($\varphi_{11}=\varphi_{22}$ and $\varphi_{12}=-\varphi_{21}$) reduce the dimension of $span(S)$ from $4$ to $2$. On the other hand, for $n>2$, one shows that $SO(n)$ matrices span the whole space of $n\times n$ real matrices \cite{son_dim}, \ie $span(S)=E$ and $d_{SO(n)}=n^2=N$.

\subsection{Convexity of the cumulant generating function and existence of the effective action}

\subsubsection{Convexity of the cumulant generating function}

Having defined the dimensionality of the microscopic field, as well as a proper basis for the span of $S$, we can address the convexity of the partition function $W[\bJ]$. 
There are two general cases: 
\begin{itemize}
\item[i)]If $d_S=N$, we show below that $W[\bJ]$ is necessarily a strictly convex function, using a textbook argument; 
\item[ii)] If $d_S<N$ due to the (effectively) linear constraints, then $W[\bJ]$ cannot be strictly convex if $\bJ\in E$. Indeed, we can decompose $\bJ$ in an orthonormal basis of $span(S)$ and $span(S)^\perp$, $\bJ=J^\parallel_i  \b{f}^\parallel_i+J^\perp_i  \b{f}_i^\perp$. Then $\bJ.\bvphi=J^{\parallel}_i\varphi_i$ does not involve $\{J^\perp_i\}_{i=1,\ldots,m}$, and $W[\bJ]$ is in fact a function of the $d_S$ variables $\{J_i^\parallel\}_{i=1,\ldots,d_S}$ only. The partition function is thus flat in the directions spanned by  $\{\b{f}_i^\perp\}_{i=1,\ldots,m}$, and cannot be strictly convex. However, if $\bJ\in span(S)$ only (and is thus parametrized by only $d_S=N-m$ parameters), then the same reasoning as the one in i) can be used to show that the corresponding cumulant generating function is  strictly convex.
\end{itemize}
Let us prove i) and ii), assuming that both $\bvphi$ and $\bJ$ are in $span(S)$ and are thus of the same dimension $d_S$ (in the first case $span(S)=E$). For this, it is sufficient to show that the Hessian of $W[\bJ]$ (the correlation function) is strictly positive for all $\bJ\in span(S)$, \ie with $\bJ=J^\parallel_i  \b{f}^\parallel_i$. Decomposing the field in the same basis, $\bvphi=\varphi_i  \b{f}^\parallel_i$, we have
\begin{equation}
G_{ij}[\bJ]=\frac{\delta^2 W}{\delta J^{\parallel}_i\delta J^{\parallel}_j}[\bJ]=\langle(\varphi_i-\langle \varphi_i \rangle_{\bJ})(\varphi_j-\langle \varphi_j \rangle_{\bJ})\rangle_{\bJ}. 
\end{equation}
Let us assume that there exist for some $\bJ$ an eigenvector $\b{u}\in span(S)$ of $G_{ij}[\bJ]$ with zero eigenvalue, which would imply that $ u_i G_{ij}[\bJ]u_j=0$. But this is equivalent to $\langle\left[\b{u}.(\bvphi-\langle\bvphi\rangle_{\bJ})\right]^2\rangle_{\bJ}=0$. Since the integration measure and $\exp(\bvphi.\bJ)$ are both strictly positive, this implies that $\b{u}.(\bvphi-\langle\bvphi\rangle_{\bJ})=0$ for all $\bvphi \in S$. But this is a linear constraint on $\bvphi$, in contradiction with the fact that the $\bvphi$'s form a (in practice overcomplete) basis of $span(S)$. Therefore $G_{ij}$ is a strictly positive matrix and $W[\bJ]$ is strictly convex if $\bJ \in span(S)$.

\subsubsection{Existence of the effective action \label{sec_TL}}

The conditions for a strictly convex generating function having been discussed, we can finally address its Legendre transform, and the existence of the effective action. For this, one needs to keep in mind that $W$ depends on $d_S\leq N$ variables, called $\{J_i^\parallel\}_{i=1,\ldots,d_S}$ above. Therefore, we only have $d_S$ components of conjugated magnetization $\{\phi_i^\parallel\}_{i=1,\ldots,d_S}$, which will be the proper variables on which the effective action will depend. The Legendre transform with respect to $\{J_i^\parallel\}_{i=1,\ldots,d_S}$ thus reads
\begin{equation}
\begin{split}
\Gamma[\bphi^\parallel]&=-W\left[\bJ[\bphi^\parallel]\right]+ \phi_i^\parallel\, J^{\parallel}_i[\bphi^\parallel] ,\\
\frac{\delta\Gamma}{\delta \phi_i^\parallel}&=J^{\parallel}_i.
\end{split}
\end{equation}
We have used the notation $\bphi^\parallel\equiv \phi_i^\parallel \b{f}_i^\parallel$ to remind the reader that the magnetization belongs to (a subset) of $span(S)$ only, and not to $E$, the space of the unconstrained microscopic fields. While $W[\bJ]$ can be defined for arbitrary source in $E$ (even though it might not be strictly convex), this is not possible for $\Gamma$, because it is constructed as a Legendre transform. 

This does not preclude the reconstruction of all correlation functions of $\bvphi$ from the effective action. For instance, in the case of the correlation function $G[\bJ]=\langle \bvphi \otimes \bvphi \rangle_{\bJ}-\langle \bvphi\rangle_{\bJ} \otimes \langle\bvphi \rangle_{\bJ}$, we have that
\begin{equation}
\begin{split}
G[\bJ]&=\langle(\varphi_i-\langle \varphi_i \rangle_{\bJ})(\varphi_j-\langle \varphi_j \rangle_{\bJ})\rangle_{\bJ} \b f_i^\parallel\otimes\b f_j^\parallel,\\
&=\frac{\delta^2 W[\bJ]}{\delta J^{\parallel}_i\delta J^{\parallel}_j}\b f_i^\parallel\otimes\b f_j^\parallel.
\end{split}
\label{eq_recovG}
\end{equation}
Since from the definition of the Legendre transform one obtains that
\begin{equation}
\sum_k \frac{\delta^2\Gamma}{\delta \phi_i^\parallel\delta \phi_k^\parallel}[\bphi]\frac{\delta^2 W}{\delta J^{\parallel}_k\delta J^{\parallel}_j}[\bJ]=\delta_{ij},
\label{eq_GGamma2}
\end{equation}
one can reconstruct the correlation function from the knowledge of $\Gamma[\bphi]$. In fact, $\frac{\delta^2\Gamma}{\delta \phi_i^\parallel\delta \phi_j^\parallel}\b f_i^\parallel\otimes\b f_j^\parallel$ is nothing but the Moore-Penrose pseudo-inverse of $G[\bJ]$ (evaluated at $\bJ[\bphi]$).

We would also like to comment on the range of definition of the effective action. Indeed, while  the magnetization  is in the span of $S$, it will usually only be in a subset of $span(S)$ that we call  $S^*=\{\bphi\; | \;\bphi=\langle \bvphi\rangle_{\bJ}, \quad \forall \bJ \in E\}$. For example, in the $XY$ toy-model, $S$ is the unit circle, and the magnetization is always of norm smaller than one, $|\bphi[\bJ]|\leq 1$, and   $S^*$ is the unit disk, even though $span(S)=\mathbb{R}^2$.\footnote{In fact, by Jensen's inequality, one shows that if the (properly defined) norm of the the microscopic field is bounded because of the constraints, $|\bvphi|\leq F$ for some $F$, then $|\bphi[\bJ]|\leq F$.}
From the theory of Legendre transform \cite{Touchette}, one can show that the fact that  $\bphi[\bJ]$ is bounded is related to the fact that $W[\bJ]$ is asymptotically a linear function of the source for large source. (What bounded, large and being linear mean in practice depends on the particular model at hand, see the examples below.) In particular, this linear behavior is also related to the rigidity of the magnetization at large source. Indeed, since the macroscopic field is constrained, increasing the source when it is already large won't change much the magnetization, implying that the correlation function $G_{ij}$ (which can be thought of as a response function here) must vanish as the source increases.
From Eq.~\eqref{eq_GGamma2}, we observe that it implies that the two-point function $\frac{\delta^2\Gamma}{\delta \phi_i^\parallel\delta \phi_k^\parallel}$ must diverge as $\bphi$ reaches its maximum value. This divergence is physical, and is a direct consequence of the rigidity of the magnetization at large source.

In summary, the condition for the cumulant generating function to be strictly convex, and the effective action to exist, is that the source belongs to the space spanned by the microscopic field. If this space is not the whole space of the unconstrained theory, then the number of independent components of the source is smaller than one might have naively thought.
These different aspects can be exemplified with the toy-models. 

\subsubsection{Application to the toy-models \label{sssec_toyTL}}

\paragraph{Symmetric matrices-}
We have seen that for symmetric $2\times2$ real matrices, $W[\bJ]$ given in Eq.~\eqref{eq_defW_sym} is not strictly convex, since it does not depend on the antisymmetric part of $\bJ$. Choosing a symmetric source to obtain strict convexity,
\begin{equation}
\bJ_s=\begin{pmatrix}
\a & \g \\
\g &\be
\end{pmatrix},
\end{equation}
one gets
\begin{equation}
\begin{split}
W[\a,\be,\g]&=\frac{1}{2}\left(\a^2+\be^2+2\g^2\right),\\
&=\frac{1}{2}{\rm tr}(\bJ_s^2).
\end{split}
\end{equation}
It is straightforward to show that the Hessian of $W[\a,\be,\g]$ is positive definite, proving strict convexity.
Note that while we have written $W$ as a function of a (symmetric) matrix $\bJ_s$, it is a function of three variables only. In particular, the derivatives can only be performed with respect to $\a$, $\be$, and $\g$, and not the four matrix elements of the source (since they are not all independent).

Calling $A$, $B$, and $\tilde C$ the conjugated variables to $\a$, $\be$, and $\g$ respectively (note that $\tilde C=2\langle c\rangle$, with $c$ the off-diagonal element of $\bvphi$, see Eq.~\eqref{eq_sym_vphi}), one obtains the effective action
\begin{equation}
\Gamma[A,B,\tilde C]=\frac{1}{2}\left(A^2+B^2+\frac{\tilde C^2}{2}\right).
\end{equation}
Collecting $A$, $B$, and $\tilde C$ in the symmetric matrix 
\begin{equation}
\tilde\bphi_s=\begin{pmatrix}
A & \frac {\tilde C}2 \\
\frac {\tilde C}2 & B
\end{pmatrix},
\end{equation}
one can write the effective action as an explicitly invariant functional $\Gamma[A,B,\tilde C]=\frac{1}{2}{\rm tr}(\tilde\bphi_s^2)$, but as for $W$, its derivatives are to be performed with respect to $A$, $B$, and $\tilde C$, and not with respect to the four matrix elements of $\bphi_s$.

\paragraph{$XY$ model-} From the results of Sec.~\ref{sssec_XY}, we have for the $XY$ model ($\bJ$ is a two-dimensional vector of length $J$)
\begin{equation}
\begin{split}
W[\bJ]&=w(J)\equiv\ln I_0(J),\\
\bphi[\bJ]&=\frac{\bJ}{J}w'(J),\\
G_{ij}[\bJ]&=\left(\delta_{ij}-\frac{J_i J_j}{J^2}\right)\frac{w'(J)}{J}+\frac{J_i J_j}{J^2}w''(J).
\end{split}
\end{equation}
Since $\det G=\frac{w'(J) w''(J)}{J}>0$, $W[\bJ]$ is strictly convex as expected.

Using the asymptotic expansion of the Bessel function at large argument, one finds that $\bphi[\bJ]=\frac{\bJ}{J}$ for sources with large amplitude, implying $\phi[J]\equiv |\bphi[\bJ]|\to 1$ as $J\to \infty$. Physically, this behavior is due to the fact that the spin is completely polarized in the direction of $\bJ$ for large field, which translates into the fact that $W[\bJ]=J$  in this same limit. This allows to obtain $G_{ij}=\frac{1}{J}\left(\delta_{ij}-\frac{J_i J_j}{J^2}\right)$ at large source, which indeed vanishes in the limit $J\to \infty$ (this is to be contrasted with $G_{ij}=\frac{\delta_{ij}}{2}$ for small sources).

Using the invariance under rotation of $W[\bJ]$, one shows that the source can be written as $\bJ_{XY} = \frac{\bphi}{\phi} F(\phi)$, with $\phi=|\bphi|$ and $F(x)$ is the inverse function of $w'(x)=\frac{I_1(x)}{I_0(x)}$. The effective action is thus invariant under rotation of the magnetization, and is given implicitly by 
\begin{equation}
\Gamma[\bphi]=-\ln I_0(F(\phi))+\phi\, F(\phi).
\end{equation}
For magnetization with amplitude close to 1, one finds
\begin{equation}
\Gamma[\bphi]=\frac{1}{2}\ln\left(\frac{\pi}{1-\phi}\right)-\frac12,
\end{equation}
which diverges on the boundary of $S^*$, \ie on the unit circle. Its derivatives also diverge, reflecting the rigidity of the magnetization, and in particular the fact that even infinitely strong sources cannot impose a magnetization $\phi>1$.

\paragraph{$SO(2)$ matrix model-} The magnetization of the $SO(2)$ matrix model is of dimension $d_{SO(2)}=2$ due to the effectively linear constraints. The two basis elements of $S=span(SO(2))$ are the matrices
\begin{equation}
\begin{split}
\b f_1^\parallel&=\frac{1}{\sqrt{2}}\begin{pmatrix}
1 &0 \\ 0 & 1
\end{pmatrix},\\
\b f_2^\parallel&=\frac{1}{\sqrt{2}}\begin{pmatrix}
0 &1 \\ -1 & 0
\end{pmatrix}.
\end{split}
\end{equation}
We need to impose that $\bJ\in span(SO(2))$, \ie $\bJ= J_d  \b f_1^\parallel+J_a \b f_2^\parallel$.
The generating function of the $SO(2)$ matrix model is then similar to that of the $XY$ model, $W[\bJ]=\ln I_0(\sqrt{2(J_a^2+J_d^2)})$, and is thus strictly convex in this subspace. The magnetization can be written as $\bphi= \phi_d  \b f_1^\parallel+\phi_a \b f_2^\parallel$, and  since we have two independent components for both the magnetization and the source, this relationship can be inverted and the Legendre transform can be performed. This would not be the case if we had kept four independent components of the source (as we would still have only two independent components for the magnetization). 

To perform the Legendre transform, it is convenient to write parametrize the source and the magnetization as $\bJ=\frac J2 O_J$ and $\bphi=\phi \,O_\phi$, with $O_J$ and $O_\phi$ in $SO(2)$, then one shows that $\bJ[\bphi]=\frac{F(\phi)}{2} O_\phi $ where $F(x)$ is the same function that the one introduced for the $XY$ model. The effective action then also takes the same form,
\begin{equation}
\Gamma[\bphi]=-\ln I_0(F(\phi))+\phi\, F(\phi).
\end{equation}
Note that $\Gamma_{SO(2)}[\bphi]$ is a function of $\phi_d$ and $\phi_a$  only (and due to the $SO(2)$ invariance, of the combination $\sqrt{\phi_d^2+\phi_a^2}=\frac{\phi}{\sqrt{2}}$), and not really a function of a $2\times2$ matrix.

\paragraph{$SO(3)$ matrix model-} The effective action of the $SO(3)$ matrix model cannot be computed explicitly, since we do not have an expression for the generating function. However,  it can be checked explicitly for small $\bJ$, and numerically for arbitrary $\bJ$, that the correlation function is indeed always strictly positive definite. 

One can compute its small magnetization expansion
\begin{equation}
\Gamma[\bphi]=\frac{3}{2}{\rm tr }(\bphi \bphi^T)+\mathcal O(\bphi^3),
\end{equation}
where $\bphi$ is an arbitrary (but small) $3\times3$ matrix. In the large source limit, one shows that $\bphi[\bJ]=\b{U}\b{V}^T$ where $\b{U},\b{V}\in SO(3)$ are such that $\b{U}^T\bJ\b{V}$ is diagonal (they are related to the singular value decomposition of $\bJ$). Similarly, one shows that the correlation function vanishes in this limit. 

Here we obtain a result which goes against naive expectations: while $SO(3)$ matrices have only three non-linear degrees of freedom, the corresponding magnetization can be viewed as an arbitrary $3 \times 3$ matrix, \ie it has $9$ independent components. This can be generalized $SO(n)$ with $n>2$, because $span(SO(n))=\mathcal{M}_n(\mathbb{R})$ (and in fact to $O(n)$ for all $n>1$ since $span(O(n))=\mathcal{M}_n(\mathbb{R})$).

\section{Systematic  treatment of  linear constraints using constrained derivatives \label{sec_derivative}}

In the previous sections, we have formally constructed a solution to the problem of defining the effective action of a constrained system using a source and a magnetization in $span(S)$. For example, for real symmetric matrices $\bvphi=\bvphi^T$, one obvious choice is to only work  with the upper triangular part of $\bvphi$. One drawback is that the transformations of the magnetization and other correlation functions under the symmetry of the partition function might not be explicit anymore, \ie the transformation $\bvphi\to U\bvphi$, which implies $\bphi[\bJ]\to U \bphi[U^{-1}\bJ]$, might not have the same simple form when written for $\bphi^\parallel$. This can make the analysis of the problem at hand cumbersome by obfuscating the original symmetries of the problem. Furthermore, one is usually interested in the correlation functions acting on the full space $E\otimes E$ (\ie $\langle \bvphi\otimes \bvphi\rangle$, as obtained from derivatives with respect to an unconstrained source). While these can be recovered using Eq.~\eqref{eq_recovG}, this can be once again cumbersome to deal with. 

It would thus  be convenient to work with functionals that are explicitly invariant, \ie that we could consider as effectively functionals of unconstrained sources or magnetization, and not just of linearly independent DOF, while still taking the linear constraints into account to allow for the existence of the effective action. The solution to this problem is known for symmetric matrices, so we start with this example before generalizing it.

\subsubsection{A detour by the case of symmetric matrices }

With the notations of Sec.~\ref{subsec_sym}, we have $W[\a,\be,\g]=\frac{1}{2}{\rm tr}(\bJ_s^2)$, which is explicitly invariant, but note that one only perform derivatives with respect to $\a$, $\be$, and $\g$, the linear DOF of the source. And while the covariance of $\langle \bvphi\rangle_{\bJ}$ under rotation is explicit, $\b V\langle \bvphi\rangle_{\b V^T\bJ\b V}\b V^T$, this is not the case of the derivatives of $W$ with respect to $\a$, $\be$, and $\g$, \ie the transformation of $\phi^\parallel_i[\bJ_s]$ is not as explicitly covariant.

In this case, the trick of how to respect the constraints of the model while still treating all components of the sources as independent is well-known, and consists in changing the definition of the derivatives. That is, instead of the standard matrix derivative of the matrix element $J_{\mu \nu}$ with respect to $J_{\rho \sigma}$,
\begin{equation}
\frac{\delta J_{\mu \nu}}{\delta J_{\rho \sigma}}=\delta_{\mu \rho}\delta_{\nu \sigma},
\end{equation}
one uses the ``constrained derivative''
\begin{equation}
\frac{\delta_c J_{\mu \nu}}{\delta J_{\rho \sigma}}\equiv\frac{\delta_{\mu \rho}\delta_{\nu \sigma}+\delta_{\nu \rho}\delta_{\mu \sigma}}2=P_{\mu \nu,\rho \sigma}.
\label{eq_der_sym}
\end{equation}
Here, $P_{\mu \nu,\rho \sigma}$ is the projector onto the subspace of symmetric matrices, \ie the projector onto $span(S)$: $P_{\mu \nu,\rho \sigma}M_{\rho \sigma}=M_{\mu \nu}$ if $\b M$ is symmetric, and $P_{\mu \nu,\rho \sigma}M_{\rho \sigma}=0$ if it is antisymmetric. In effect, one is performing the derivative as if all matrix elements were independent,  projecting onto the subspace of symmetric matrices, and then assuming that the matrix is symmetric
\begin{equation}
\frac{\delta_c W[\bJ_s]}{\delta J_{\mu \nu}}=P_{\mu \nu,\rho \sigma}\frac{\delta W[\bJ]}{\delta J_{\rho \sigma}}\bigg|_{\bJ=\bJ_s}.
\end{equation}
No information is lost when applying the projector since $\frac{\delta W[\bJ]}{\delta \bJ}\in span(S)$ for any source (symmetric or not), and using the constrained derivative on $W[\bJ_s]$  is equivalent to a standard derivative with respect to  $\bJ\in E$ (evaluated in $\bJ=\bJ_s$),
\begin{equation}
\frac{\delta_c W[\bJ_s]}{\delta \bJ}=\frac{\delta W[\bJ]}{\delta \bJ}\bigg|_{\bJ=\bJ_s}.
\end{equation}
And while $\frac{\delta W[\bJ_s]}{\delta \bJ}$ is meaningless, $\frac{\delta_c W[\bJ_s]}{\delta \bJ}$ gives a well-defined and useful definition of the derivative of a symmetric matrix with respect to all its matrix elements. Importantly, there is not any ambiguity for the derivative of the off-diagonal elements. Indeed, even though we can write $\gamma= a_1 J_{12}+a_2 J_{21}$ with any $a_1$ and $a_2$ such that $a_1+a_2=1$, its constrained derivative is always the same,
\begin{equation}
\frac{\delta_c \gamma}{\d J_{12}}=\frac{1}{2}=\frac{\delta_c \gamma}{\d J_{21}}.
\end{equation}
Thus the constrained derivative of $W$ reads
 \begin{equation}
 \frac{\delta_c W[\bJ_s]}{\delta \bJ}=\begin{pmatrix}
 \alpha & \gamma \\
 \gamma & \beta
 \end{pmatrix}=\begin{pmatrix}
 \langle a\rangle & \langle c\rangle\\
 \langle c\rangle & \langle b\rangle
 \end{pmatrix}=\bphi[\bJ_s].
 \end{equation}
Note that $ \frac{\delta_c W[\bJ_s]}{\delta \bJ}$ (and all higher order derivatives) now transform nicely under the symmetry transformations. Furthermore, introducing the magnetization as a symmetric matrix
\begin{equation}
\bphi_s=\begin{pmatrix}
A &   C \\
  C & B
\end{pmatrix},
\end{equation}
the inversion between $\bJ_s$ and $\bphi_s$ can be performed easily, and one is lead naturally to the effective action
\begin{equation}
\Gamma[\bphi_s]=\frac{1}{2}{\rm tr}\left(\bphi_s^2\right).
\end{equation}
Therefore, using this constrained derivative, which in effect projects all quantities onto $span(S)$ to which $\bvphi$ and $\bphi$ belong, one can work as if there were no constraints on the fields when performing derivatives, with respect to the source or the magnetization. This additionally has an advantage of making the magnetization and higher correlation functions explicitly covariant under the symmetries of the problem.

\subsubsection{Generalization to arbitrary linear constraints}
Having understood how to solve the problem of the constraints for the simple case of symmetric matrices, we are now in position to generalize this strategy to $m$ arbitrary linear constraints $\{L_a\}_{a=1,\ldots,m}$ (possibly induced by the non-linear constraints).

A general method to devise constrained differentiation has been devised by Schay \cite{Schay1995}, which we summarize now.
 If we perform an arbitrary increment $\delta\bJ$ of $\bJ$, not necessarily satisfying the constraints, we can project these increments on the constrained subspace: $\delta_c \bJ=P\delta\bJ$, with the projector $P_{AB}(\bJ)=\delta_{AB}-\frac{\delta L_a(\bJ)}{\delta J_A} H_{ab}\frac{\delta L_b(\bJ)}{\delta J_B}$ with $H_{ab} \frac{\delta L_b(\bJ)}{\delta J_A}\frac{\delta L_c(\bJ)}{\delta J_A}=\delta_{ac}$.  Note that since we only need to consider the linear constraints here, the matrix $H$ and the projector $P$ do not depend at what point  the derivative is performed. 
 Then the increment of a function $f$ is given by 
 \begin{equation}
 \delta_c f(\bJ)=\sum_{A,B=1}^N\frac{\delta f(\bJ)}{\delta J_A} P_{AB} \delta J_B,
 \end{equation}
that is,
\begin{equation}
 \frac{\delta_c J_A}{\delta J_B}\equiv P_{AB},
\end{equation}
which is our constrained derivative. In the present case, the constraint is that $\bJ\in  span(S)$. Since we already have a basis for this space, 
$\{\b{f}^\parallel_i\}_{i=1,\ldots, d_S}$, the projector is trivial to write down,
\begin{equation}
P_{AB}=f^{\parallel}_{i,A} f^{\parallel}_{i,B},
\end{equation}
with $f^{\parallel}_{i,A}=\b e_A.\b f^{\parallel}_i$. 
 One shows easily that for symmetric matrices, one recovers Eq.~\eqref{eq_der_sym}.
Since both $\bJ$ and $\bphi$ are in $span(S)$, the constrained derivative is the same for both fields. (This is not necessarily the case for a more general bilinear coupling between source and field, see Appendix \ref{app_general}.)

With this definition, the Legendre transform is performed as follows. Starting from $W[\bJ]$ with $\bJ\in span(S)$, one obtains
\begin{equation}
\phi_A[\bJ]=\frac{\delta_c W[\bJ]}{\delta J_A}.
\end{equation}
Since  $\bphi$ and $\bJ$ are in $span(S)$, we can invert this relationship to obtain $\bJ[\bphi]$, and the effective action is given by
\begin{equation}
\Gamma[\bphi]=-W[\bJ[\bphi]]+\bphi.\bJ[\bphi],
\end{equation}
and is well defined. The effective action obtained that way has the same form than the one obtained with the more conventional method of Sec.~\ref{sec_TL}. But while the two methods are equivalent at this level, the advantage of the method presented here is that one is effectively working as if the fields are unconstrained, the only modification coming from the definition of the derivatives.
 For example, one shows that Eq.~(\ref{eq_GGamma2}) now becomes 
\begin{equation}
\label{eq_pseudoidentity}
\frac{\delta_c^2 \Gamma[\bphi]}{\delta \phi_A\delta \phi_B}\frac{\delta_c^2 W[\bJ]}{\delta  J_B\delta J_C}=P_{AC}.
\end{equation}
The second derivative of the effective action featured in Eq.~(\ref{eq_pseudoidentity}) is in fact the Moore-Penrose pseudo-inverse of the correlation function, \ie the inverse restricted to the subspace $span(S)$.

 \subsubsection{Application to the $SO(2)$ matrix model}
The method explained above can be exemplified with the $SO(2)$ matrix model
The constrained derivative for a $2\times2$ matrix $\bJ\in span(SO(2))$, \ie such that $J_{11}=J_{22}$ and $J_{12}=-J_{21}$ is given by
 \begin{equation}
 \frac{\delta_c J_{\mu \nu}}{\delta J_{\rho \sigma}}\equiv\frac{\delta_{\mu \rho}\delta_{\nu \sigma}-\delta_{\nu \rho}\delta_{\mu \sigma}+\delta_{\mu \nu}\delta_{\rho \sigma}}2=P_{\mu \nu,\rho \sigma}.
 \label{eq_dJdJ}
 \end{equation}
 Starting from the $W[\bJ]=\ln I_0(f(\bJ))$ with $f(\bJ)=2\sqrt{\det \bJ}$ for $\bJ\in span(SO(2))$, and using that $\frac{\delta_c f(\bJ)}{\delta J_{\mu\nu}}=\frac{2}{f(\bJ)}J_{\mu\nu}$, one gets 
 \begin{equation}
 \bphi[\bJ]=\frac{2\bJ}{f(\bJ)}\frac{I_1\left(f(\bJ)\right)}{I_0\left(f(\bJ)\right)},
 \end{equation}
 which is of course also in $span(SO(2))$. This relationship between $\bJ$ and $\bphi$ can be inverted, and we find $\bJ[\bphi]=\frac{\bphi}{2\phi}F(\phi)$, where $\phi=\sqrt{\det \bphi}$ and $F(x)$ is the same function than defined in Sec.~\ref{sec_TL}. The Legendre transform is well defined, and given by
 \begin{equation}
 \begin{split}
 \Gamma[\bphi]&=-W[\bJ[\bphi]]+\Tr\left(\bphi^T.\bJ[\bphi]\right),\\
 &=-\ln I_0(F(\phi))+\phi\, F(\phi),
 \end{split}
 \end{equation}
 which has the same form than that obtain in Sec.~\ref{sssec_toyTL}.

\section{Discussion \label{sec_concl}}

We have shown that to properly define the effective action, non-linear constraints do not pose any problem as long as they do not imply linear constraints between the components of the microscopic field. In contrast, linear constraints (possibly induced by non-linear ones) reduce the number of linear DOF the microscopic field. For the generating function of the cumulant to be strictly convex, the source must only couple to these linear DOF, which then allows for obtaining a well-defined effective action. Finally, the magnetization has the same number of DOF than the effectively linearly independent DOF of the microscopic field, \ie it belongs to a subset of the space spanned by the microscopic field. 

We are now in a position to comment on what should be considered the ``real'' DOF in our opinion. The effective action is a much more physically meaningful object than the Hamiltonian or action, since it includes both  constraints and  fluctuations. As such, it is natural to consider the magnetization as the more physical fields, the DOF of which should be viewed as the ``real'' ones. And since the magnetization only respects the (effectively) linear constraints of the microscopic field (the non-linear constraints only change its range of definition), we should consider as ``real'' the linear DOF of the microscopic field, and not the non-linear ones. This leads to the rather counter-intuitive fact that for the $SO(n)$ matrix models with $n>2$, where the number of non-linear constraints $M$ is larger than half the number $N$ of components of the microscopic field, the true number of DOF is in fact $N$: all components of the magnetization are independent.

A very interesting question is to understand to what extent the formalism developed here can be applied to other field theories, such as quantum and supersymmetric models. In the quantum case, the existence of a locally conserved field (for instance, the spin component along the source/magnetic field) can preclude the existence of a local effective action. On the other hand, the Grassmann variables of supersymmetric models can render the measure of the functional integral negative, forbidding the reasoning used here to insure the strict convexity of cumulant generating functions. In fact, the partition function can even vanish (and then become negative) for finite values of the source, which poses a number of problems to define a proper effective action. While perturbation theory is insensitive to these issues (as it is formulated for infinitesimal values of the sources), circumventing these problems is necessary to formulate a fully functional version of the FRG and to go beyond perturbative calculations. This would be beneficial for non-perturbative study of dense loop soup models \cite{Jacobsen2003} and Anderson localization \cite{Efetov1983}, that both rely on supersymmetric models. We leave these intriguing problems for future work.

\subsection*{ACKNOWLEDGMENTS}
A.R. warmly thanks G. Verley, N. Dupuis and B. Delamotte for numerous discussions, as well as H. Jacquin for a careful reading of the manuscript. A.R. is indebted to  the attendees of the ERG 2018 conference for discussions, as well as the MathOverflow community. A.R. thanks the Institute of Physics of Zagreb where part of this work was done.  I.B. acknowledges the support of the Croatian Science Foundation Project No. IP-2016-6-7258 and the QuantiXLie Centre of Excellence, a project cofinanced by the Croatian Government and European Union through the European Regional Development Fund - the Competitiveness and Cohesion Operational Programme (Grant KK.01.1.1.01.0004). A.R. acknowledges the support of the Programme Investissement d'Avenir (I-SITE ULNE / ANR-16-IDEX-0004 ULNE).

 \appendix

 \section{More general coupling between sources and fields \label{app_general}}

 We generalize the results of the main text to the case where the coupling between the source $\bJ$ and the microscopic field $\bvphi$ is not necessarily a definite positive bilinear form. We write this coupling as $(\bJ,\bvphi)=J_A \eta_{AB}\varphi_B$ with $\eta_{AB}=(e_A,e_B)$ not necessarily definite positive. Our assumptions are that $\eta$ is symmetric ($\eta_{AB}=\eta_{BA}$), non-degenerate, and to alleviate the notations, that it is its own inverse ($\eta_{AB}\eta_{BC}=\delta_{AC}$).

 This case is inspired from a supersymmetric non-linear sigma model \cite{RanconBalog}, which for our present purpose can be simplified into a vector model $\bvphi=(\varphi_1,\varphi_2,\varphi_3,\varphi_4)$ with linear constraint $\varphi_1-\varphi_2+\varphi_3-\varphi_4=0$ and $\eta={\rm diag}(1,-1,1,-1)$.

  The cumulant generating function is now given by
 \begin{equation}
 W[\bJ]=\ln \int d\mu(\bvphi)\, e^{(\bJ,\bvphi)},
 \end{equation}
 and the magnetization reads
 \begin{equation}
 \phi_A=\eta_{AB}\frac{\delta W[\bJ]}{\delta J_B}.
 \end{equation}

 As in Sec.~\ref{sec_existenceTL}, the dimension of $\bvphi$ is still $d_S=N-m$, and its span  is still $span(S)$, with basis $\{\b f_i^\parallel\}_{i=1,\ldots,d_S}$, and orthogonal complement $S^\perp$ with basis $\{\b f_i^\perp\}_{i=1,\ldots,m}$ (note that we still have the natural scalar product such that $\b e_A.\b e_B=\delta_{AB}$). However, the space to which the source must belong for $W[\bJ]$ to be strictly convex is now different. Indeed, we must have $\bJ\in \tilde S$, with $\tilde S=\{\bJ \in E \; | \; (\bJ,\b f_i^\perp)=0, \; i=1,\ldots,m\}$. Since $\eta$ is full rank, we see that a basis of  $\tilde S$ is $\{\b{\tilde{f}}_i^\parallel\}_{i=1,\ldots,d_S}$ with $\tilde{f}_{i,A}^{\parallel}=\eta_{AB} f_{i,B}^{\parallel}$. Note that $\tilde S$ is of the same dimension than $S$. One can show that  $\{\b{\tilde{f}}_i^\perp\}_{i=1,\ldots,m}$ with $\tilde{f}_{i,A}^{\perp}=\eta_{AB} f_{i,B}^{\perp}$ is a basis of $\{\bJ \in E\; | \; (\bJ,\b f_i^\parallel)=0, \; i=1,\ldots,d_S\}$, independent of that of $S$.

 Decomposing the source in the basis of $\tilde S$, $\bJ=\tilde J_i^\parallel\b{\tilde{f}}_i^{\parallel}$, the same arguments than that of Sec.~\ref{sec_existenceTL} show that $W[\bJ]$ is a strictly convex function of $\{\tilde J_i^\parallel\}_{i=1,\ldots,d_S}$. In particular, with the microscopic field written as $\bvphi= \varphi_i\b f_i^\parallel$, one finds that $(\bJ,\bvphi)=\tilde J_i^\parallel \varphi_i$, and the effective action reads
 \begin{equation}
 \Gamma[\bphi^\parallel]=-W[\bJ[\bphi^\parallel]]+\tilde J_i^\parallel[\bphi^\parallel] \phi_i,
 \end{equation}
 with $\bphi=\b f_i^\parallel \phi_i$.

 It is also useful to work with constrained derivatives, which allows for functionals that are explicitly invariants under transformations that preserve the bilinear form. The main difference is that now, the constraints on $\bJ$ and $\bphi$ are not the same, since $\bJ\in \tilde S$ and $\bphi\in span(S)$. There are no difficulties to show that 
 \begin{equation}
 \begin{split}
 \frac{\delta_c J_A}{\delta J_B}&=\tilde P_{AB}\equiv \tilde  f^{\parallel}_{i,A}\tilde  f^{\parallel}_{i,B},\\
 \frac{\delta_c \phi_A}{\delta \phi_B}&= P_{AB}\equiv  f^{\parallel}_{i,A} f^{\parallel}_{i,B},
 \end{split}
 \end{equation}
 and in particular $\tilde P_{AB}=\eta_{AC}P_{CD}\eta_{DB}$.

\bibliography{bibLegendre,/home/adam/Dropbox/Articles/bibli_RG,/home/adam/Dropbox/Articles/bibli_QFT,/home/adam/Dropbox/Articles/bibli_bosons,/home/adam/Dropbox/Articles/bibli_disorder,/home/adam/Dropbox/Articles/bibli_diverse}

\end{document}